\documentclass{jfm}
\usepackage{graphicx}
\usepackage{caption}
\usepackage{subcaption}
\usepackage{epstopdf, epsfig}
\usepackage{amsmath}
\usepackage[dvipsnames,svgnames,x11names]{xcolor}
\usepackage[final]{changes}
\definechangesauthor[color=BrickRed]{MC}

\newcommand{\ub}{{\bf u}}

\newcommand{\Ub}{{\bf U}}
\newcommand{\Xb}{{\bf X}}

\newcommand{\xb}{{\bf x}}
\newcommand{\fb}{{\bf f}}
\newcommand{\Fb}{{\bf F}}

\newcommand{\dd}{{\mathrm d}}
\newcommand{\St}{\mathit{St}}

\shorttitle{The assembly of freely moving rigid fibers measures the flow gradient tensor}
\shortauthor{M. Cavaiola, S. Olivieri and A. Mazzino}

\title{The assembly of freely moving rigid fibers measures the flow velocity gradient tensor}

\author{Mattia Cavaiola\aff{1,}\aff{2},
 Stefano Olivieri\aff{1,}\aff{2}
 \and Andrea Mazzino\aff{1,}\aff{2}
 \corresp{\email{andrea.mazzino@unige.it}}}

\affiliation{\aff{1}Department of Civil, Chemical and Environmental Engineering (DICCA), University of Genova, Via Montallegro 1, 16145, Genova (Italy)
\aff{2}INFN, Genova Section, Via Montallegro 1, 16145, Genova (Italy)}

\begin{document}

\maketitle

\begin{abstract}
The motion of an assembly of rigid fibers is investigated for different classes of closed streamline flows, steady or time dependent, two dimensional or three dimensional. In our study, the dynamics of the fiber assembly is fully-coupled to the flow field by means of a state-of-the-art immersed boundary method. We show that, for sufficiently small Stokes times of the assembly, the whole flow gradient tensor can be accurately reconstructed by simply tracking the fiber assembly and measuring suitable fiber velocity differences evaluated at the fiber ends. Our results strongly suggest the possibility of using rigid fibers (or assembl\added{ies} of them) to perform multi-point flow measures, either in laboratory or in field: future experiments are \added{therefore} mandatory to inquire the feasibility of a new `Fiber Tracking Velocimetry' technique. 
\end{abstract}

\begin{keywords}
Authors should not enter keywords on the manuscript, as these must be chosen by the author during the online submission process and will then be added during the typesetting process (see http://journals.cambridge.org/data/\linebreak[3]relatedlink/jfm-\linebreak[3]keywords.pdf for the full list)
\end{keywords}

\section{\label{sec:intro}Introduction}

\added{Understanding the dynamical behaviour of fiber-like objects dispersed in fluid flow is a relevant issue concerning many environmental and industrial processes, such as pollutant dispersion, microfluidic processing and paper production~\mbox{\citep{duroure2019review}}.
Compared with point-like particles, the dynamics of \added{fibers or, more generally, non-spherical particles, turns out to be} more complex due to the additional degrees of freedom related to
orientation, so that active research is devoted to improve the comprehension of such fluid-structure interactions.}

\added{In the case of rigid particles with ellipsoidal shape and sufficiently small to evolve in a Stokes flow with negligible fluid inertia, an analytical expression for the fluid torque acting on the particle was originally derived by~\mbox{\citet{jeffery1922motion}}. Such result has been later generalized to other shapes and widely exploited in a variety of problems.
These include studies on the rheology of suspensions in low-Reynolds flow conditions {\citep{butler2018review}}, as well as on the dynamics of dispersed fibers in turbulent flows~\mbox{\citep{voth2017review}}.
Focusing on the latter framework,  
several investigations concerned both homogeneous isotropic~\citep{parsa2012,ni_kramel_ouellette_voth_2015,sabban_cohen_van-hout_2017} as well as wall-bounded turbulent flows~\mbox{\citep{marchioli2010,marchioli2016}}, revealing useful insights on the preferential alignment experienced by fibers and the correlation statistics between their orientation and vorticity.
Additionally,~\mbox{\citet{gustavsson2019}} recently focused on the settling of small prolate particles, while the orientation of rod-like particles in Taylor-Couette turbulence was analyzed by~\mbox{\citet{bakhuis2019}}.

Despite its significance, the approach based on Jeffery's solution is justified only if the aforementioned assumptions are satisfied. For example, these no longer apply if the fiber length is larger than the Kolmogorov flow scale (i.e., the Reynolds number at the fiber lengthscale is not sufficiently small). In fact, the dynamical behavior of fibers with length well within the inertial range of scales is far less understood and has been considered by only few recent experimental investigations~\citep{bounoua2018tumbling,kuperman2019inertial}, along with the numerical study of~\citet{do2014} who simulated rigid fibers of finite size in a turbulent channel flow.
Furthermore, for Jeffery's solution to hold strictly, it is also required that particle inertia (typically quantified by means of the Stokes number) can be neglected as well~\citep{sabban_cohen_van-hout_2017}.
}

\added{A further breakdown for the application of Jeffery's model is for the case of flexible fibers.
{The dynamics of flexible fibers in low-Reynolds number flows has been recently reviewed by~\mbox{\citet{duroure2019review}}.
For sufficiently small fibers, the typical modeling approach relies here on slender body theory~\citep{cox1970,tornberg2004}.
Using this approach, in particular, the motion of flexible fibers in cellular flows has been extensively studied by~\citet{young2007stretch,wandersman2010buckled,quennouz2015transport}, revealing the existence of flow-induced buckling instabilities that are responsible of their complex dynamics, including the possibility of a diffusive behaviour for neutrally-buoyant fibers due to such deformation.}
Moreover,~\mbox{\citet{allende2018}} recently investigated the stretching rate and buckling probability of noninertial flexible fibers in ideal turbulence.

For flexible fibers with finite size,}
some recent contributions have considered this kind of objects as the key ingredient for a novel way of flow measurement.
In particular, the possibility of using flexible fibers to quantify two-point statistics has been highlighted by~\citet{rosti2018flexible,rosti2019flowing} in the case of homogeneous isotropic turbulence by means of fully-resolved direct numerical simulation. This latter case is very far from the realm of application of Jeffery's
model for at least three main reasons. The fibers are elastic, inertial
and they do not evolve locally in a linear flow (i.e. their size is well within the inertial range of scales).
In this situation, the existence of different fiber flapping states was identified, in some of which the fiber behaves as a proxy of turbulent eddies with size comparable to the fiber length. Two-point statistical quantities, such as the velocity structure functions, were thus acquired simply by tracking the fiber end points in time.
\added{Related to this framework, significant contributions regarded the flapping instabilities of flexible filaments interacting with two-dimensional low-Reynolds flow~{\citep{shelley2011flapping}}, and similar mechanisms were explored for the purpose of passive locomotion and flow control purposes~{\citep{bagheri2012symmetry,lacis2014passive,lacis2017passive}}.
Along a similar line of reasoning,~{\citet{hejazi2019gradients}} investigated experimentally how to measure fluid velocity gradients using particles made by connections of slender deformable arms, both in the case of two-dimensional shear flow and three-dimensional turbulence.}

\added{Motivated by the evidence reported by~{\citet{rosti2018flexible,rosti2019flowing}}}, the goal of this work is to investigate similar possibilities for the case of rigid fibers in laminar flow.
\deleted{Also in the present study the Jeffery's model will not apply to our fibers for at least two reasons: they will be inertial and they will be not evolving necessarily in a linear flow.}
Rigid fibers are indeed easier to fabric than elastic ones and are good candidates for novel experimental, non-invasive techniques able to access small-scale, multi-point properties of fluid flows. The idea is to replace single particles, typically used in particle image or tracking velocimetry (PIV/PTV) to measure single-point fluid properties~\citep{adrian1991particle,hoyer2005,schanz2016shake}, by single fibers (or assemblies of them) in order to access two-point (or multi-point) properties.

\begin{figure}
\centering
\includegraphics[width=0.8\textwidth]{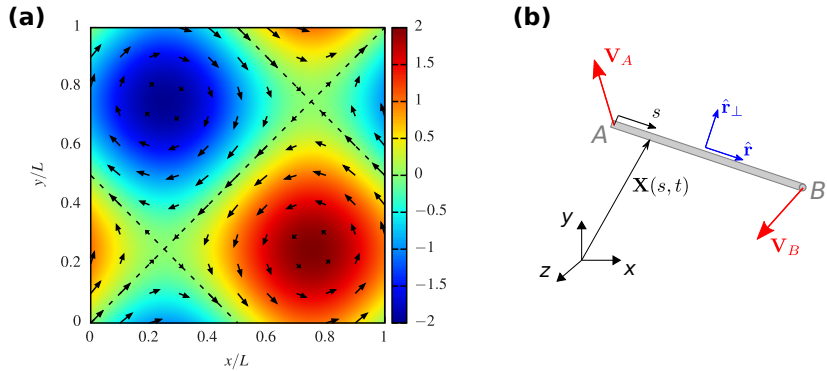}
\caption{(a) The so-called \textit{BC} cellular flow (the colormap showing the stream function given by (\ref{eq:BC-psi}) along with the corresponding velocity vectors); (b) sketch of a generic fiber configuration (the characteristic quantities here indicated are introduced in the text).}
\label{fig:sketch}
\end{figure}

To this aim, we will focus on cellular flows, which are also a conceptual representation of eddies of turbulent flows.
\deleted{The motion of flexible fibers in cellular flow has been extensively studied by~\mbox{\citet{young2007stretch,wandersman2010buckled,quennouz2015transport}} and recently reviewed by~\mbox{\citet{duroure2019review}}, revealing the existence of flow-induced buckling instabilities that are responsible of their complex dynamics, including the possibility of a diffusive behaviour for neutrally-buoyant fibers due to such deformation.}
We will \added{therefore consider} spatially-periodic solutions of the incompressible Navier-Stokes equations, i.e. the so-called \added{Arnold-Beltrami-Childress (ABC) and Beltrami-Childress (BC)} flows~\citep{dombre1986chaotic,biferale1995eddy}. A visualization of the latter is given in figure~\ref{fig:sketch}a.
The choice of this setting will enable us to perform a direct and reliable comparison between the measured fiber velocity and the underlying, unperturbed fluid flow velocity.
Although the fiber velocity is generally different from the \added{unperturbed fluid velocity}, we will show that the velocities at fiber ends can be used to measure \added{the velocity differences of  the unperturbed fluid flow. With unperturbed velocity here we mean the velocity field of the flow in the absence of the fiber}.
In this framework, a new way for measuring the fluid velocity gradient tensor will be proposed, and tested exploiting the assembly of different fibers (three fibers for the two-dimensional incompressible case
and eight fibers for the corresponding three-dimensional case).
Accessing the velocity gradient is of particular importance when dealing with turbulent flows, since from this quantity one can thus construct the vorticity and strain rate tensors, as well as obtain the energy dissipation rate and other small-scale quantities.

It is worth summarizing in which sense our model for the fiber dynamics is different with respect to the well-known  Jeffery's model.
In our model the fiber is fully-coupled to the flow and it will be of
finite size, i.e., its length will be up to the typical size of the cellular flows we will consider. In general, the fiber will be thus not evolving in a linear flow as it is in the Jeffery's model. Moreover, our fiber will be inertial. Different Stokes numbers will be analyzed with the final aim of understanding up to which Stokes numbers a rigid fiber can be used as a proxy of \added{unperturbed} flow properties.

\added{Following this introduction,} the rest of the paper is structured as follows: \S~\ref{sec:method} presents the numerical methods used in the work (complemented by Appendix~\ref{sec:appendix_IBM}), \S~\ref{sec:results} exposes the results, and finally \S~\ref{sec:discussion} draws some conclusions and perspectives.

\section{\label{sec:method}Methods}
We consider an inertial, elastic fiber of length $c$ and diameter $d \ll c$, characterized by \added{(nondimensional)} linear density $\rho_1$ and bending stiffness $\gamma$. 
Given the position of a material point belonging to the fiber $\Xb = \Xb(s,t)$, as a function of the curvilinear coordinate $s$ and time $t$, the fiber dynamics is governed by the Euler-Bernoulli's beam equation
\begin{equation}
\rho_{1}\ddot\Xb = \partial_s(T \partial_s (\Xb)) - \gamma\partial^4_{s}(\Xb) - \Fb.
\label{eq:EB}
\end{equation}
In Eq.~(\ref{eq:EB}), $\Fb$ is the forcing exerted by the fluid-structure coupling, while $T$ is the tension necessary to enforce the inextensibility condition
\begin{equation}
 \partial_{s}(\Xb) \cdot \partial_{s}(\Xb) = 1.
\end{equation}
The fiber is freely moving in the flow, hence the corresponding boundary conditions at its ends are: 
\begin{equation}
 \partial_{ss} \Xb |_{s=0,c} = \partial_{sss} \Xb |_{s=0,c} = 0,
\end{equation}
\begin{equation}
 T |_{s=0,c} = 0.
\end{equation}

Nevertheless, we focus here on rigid fibers, as sketched in figure~\ref{fig:sketch}b. To this end, throughout the work we choose and retain \added{$c/L = (2 \pi)^{-1}$ and} $\gamma = 10$ for which we have an essentially rigid behavior with negligible deformations. \added{The latter can be quantified by looking at the magnitude of the end-to-end distance, which results always smaller than $\mathcal{O}(10^{-8})$}.

The fiber is discretized along $s$ into segments with spatial resolution $\Delta s = c / (N_\mathrm{L} - 1)$, $N_\mathrm{L}$ being the number of Lagrangian points.
To model the fluid-structure coupling, we will consider two different strategies: (i) a fully-resolved approach where the feedback is taken into account (which will therefore be denoted as \emph{active})  and (ii) an intrinsically \emph{passive} model based on slender body theory.
Both strategies are introduced in the following of this section.

\subsection{Active model}
\label{sec:active}
In the first case the coupling is two-way and the dynamics is resolved using an immersed boundary (IB) technique, inspired by the method proposed by~\citet{huang_shin_sung_2007a} for anchored filaments in laminar flow. The method was also exploited for dispersed fibers in turbulent flow~\citep{rosti2018flexible,rosti2019flowing,banaei2019numerical}. In the present case, we solve numerically the incompressible Navier-Stokes equations for the fluid flow (details on the solution method can be found in Appendix~\ref{sec:appendix_IBM}):
\begin{equation}
  \partial_{t} \ub + \ub\cdot\boldsymbol{\partial}\ub = -\boldsymbol{\partial} p/\rho_0 +\nu\partial^2\ub +  \fb,
  \label{eq:NS1}
\end{equation}
\begin{equation}
  \boldsymbol{\partial}\cdot\ub = 0
  \label{eq:NS2}    
\end{equation}
where $\ub$ is the fluid velocity, $p$ the pressure, $\rho_0$ the density, and $\nu$ the kinematic viscosity.
The volumetric forcing $\fb$ is made by the sum of two contributions: the first is used for generating the desired flow field~\citep{dombre1986chaotic}, while the second is characteristic of the IB method, mimicking the presence of the fiber by means of no-slip enforcement at the Lagrangian points. 
\added{Note that the single-mode non-random nature of the volume forcing considered here to generate the flow field does not require the special attention as for the cases considered e.g. by~{\citet{chouippe2015}}.}

When using this approach, we consider a cubic domain of side $L=2\pi$ with periodic boundary conditions in all directions, which is discretized into a uniformly spaced Cartesian grid using $N=64$ cells per side.
The number of Lagrangian points describing the fiber is chosen in such a way that the Lagrangian spacing $\Delta s$ is almost equal to that of the Eulerian grid $\Delta x$: e.g., for a fiber with length $c=1$ we use $N_\mathrm{L}=11$ Lagrangian points. Doubling both resolutions, the variation of results was found to be negligible. As for the timestep we use $\Delta t = 5 \times 10^{-5}$, after assessing the convergence for this parameter as well.
\subsection{Passive model}
\label{sec:pass}
In the second approach, a one-way coupling is assumed, i.e. the fiber is forced by the flow but not vice versa. The problem thus essentially consists in solving only Eq.~(\ref{eq:EB}), \added{within which the forcing term is expressed} as
\begin{equation}
 \Fb = \frac{\rho_1}{\tau_s} \, (\dot \Xb - \ub(\Xb(s,t),t)),
 \label{eq:FP1}  
\end{equation}
\added{$\tau_s$ being the relaxation time of a fiber immersed in viscous flow, i.e. it can be identified with a Stokes time}. \deleted{Because of the isotropy of the drag force, there is no way to distinguish whether it is a rotational or a translational Stokes time.} Overall, it represents a measure of the fiber inertia compared to the flow, and also quantifies the strength of the coupling. In this approach, $\ub$ is assigned and does not get modified by the presence of the fiber.
Note that our model is similar (apart from the contribution of inertia that we have accounted for in our model) to that of~\citet{young2007stretch,quennouz2015transport}, relying on local slender body theory, and further simplified assuming an isotropic form for $\Fb$.

Using this model, we assess the convergence of the solution only with the Lagrangian resolution (since there is not the need of an Eulerian grid in this case) and the computational time step.
Testing was performed for several initial positions of the fiber and in different cellular flow configurations (i.e., 2-D or 3-D, steady or oscillating).
\added{Given the substantially lower computational demand of this approach compared with the active model,}
for a fiber with length $c=1$ \added{we choose} $N_\mathrm{L}=31$ and $\Delta t =  10^{-6}$\added{, although a numerically stable and resolution-independent solution is  already found with coarser resolution, accordingly with our findings for the active case.} \deleted{Increasing $N_\mathrm{L}$ and/or decreasing $\Delta t$, negligible variations were observed.}
\deleted{One can note that these values differ from those used in the active case and that, in particular, the requirements on both the spatial and temporal resolution are found to be more strict for the passive model. This fact can be explained since in the active approach the no-slip condition is properly enforced on the flow; consequently, in the vicinity of the fiber the fluid flow gradient gets smoothed, this relaxing the requirements for numerical stability.}

\section{Results}
\label{sec:results}
\subsection{\label{sec:BC}Two-dimensional BC flow}
To start our analysis, we consider the steady and two-dimensional cellular flow, often named as the Beltrami-Childress (BC) flow, that is defined by the stream function
\begin{equation}
    \Psi(x,y) = \sin(y) - \sin(x),
    \label{eq:BC-psi}
\end{equation}
\added{which is shown} in figure~\ref{fig:sketch}a. From Eq.~(\ref{eq:BC-psi})\added{,} it follows that the velocity field can be expressed as:
\begin{equation}
 \begin{array}{ll}
    u &= \cos y \\
    v &= \cos x
  \end{array}
  \label{eq:BC}
\end{equation}
Such a relatively simple flow configuration will first be used for assessing the importance of fiber inertia by evaluating the rotational Stokes number, and then to present the actual method for measuring two-point velocity differences.

\subsubsection{\label{sec:StNum}Rotational Stokes number}

\begin{figure} 
\centering

\includegraphics{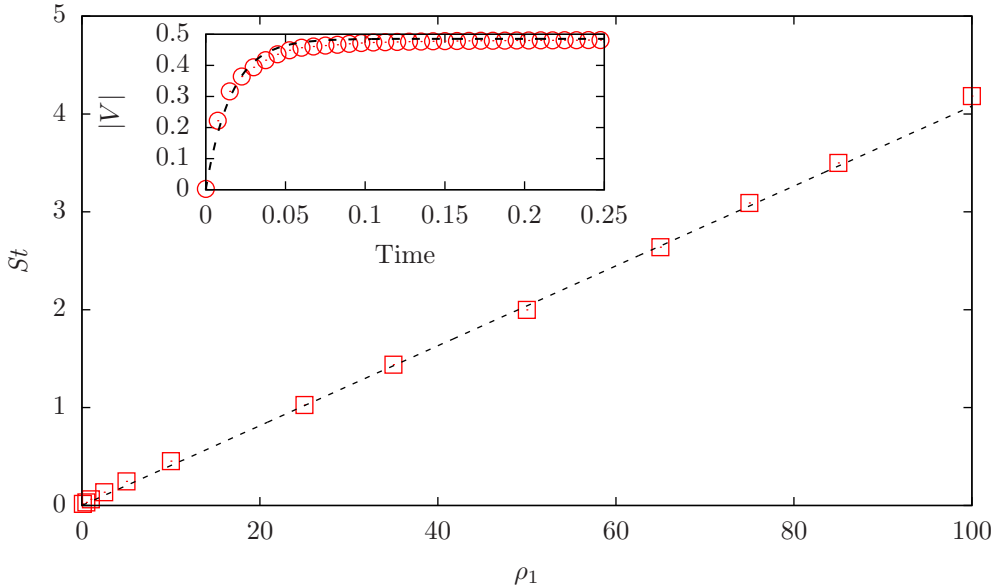}
\caption{\added{Rotational Stokes number as a function of fiber linear density. Red squares: measured values using the active model; black dashed line: linear fit. 
Inset: time history of the velocity magnitude of one fiber end (red circles)  fitted using Eq.~(\ref{eq:fit}) (black dashed line), for a fiber with $\rho_{1}$ = 0.1.}}
\label{fig:StvsRho1}
\end{figure}

As a preliminary step, we characterize the effect of fiber inertia by estimating the rotational Stokes number, which is the most suitable quantity to consider when dealing with cellular flows such as those considered in this work. \added{This has been done for the sole active fibers, the Stokes number being an assigned parameter in the passive case as shown by~Eq.~(\ref{eq:FP1}).}

To this end, we proceed as follows. We place the fiber at the center of one cell in the BC flow (figure~\ref{fig:sketch}a). The fiber is initially at rest and, under the action of the flow, will start to purely rotate around its center of mass. \replaced{W}{Hence, w}e measure the time it takes for the fiber to adapt to the flow, i.e., focusing on the velocity magnitude of one fiber\deleted{'s} end $V(t)$, to assume constant velocity $V_0$ compatible with the unperturbed fluid velocity. Hence, we perform an exponential fit
\begin{equation}
  V(t)=V_0(1-e^{-t/\tau_{s}})
  \label{eq:fit}
\end{equation}
to measure the Stokes time $\tau_{s}$. An example of this procedure is given by the inset of figure~\ref{fig:StvsRho1}.
From the best fit we obtain the Stokes time and thus the Stokes number defined as $\mathit{St} = \tau_{s} / \tau_{f}$, where $\tau_{f} = c/U$ is the characteristic hydrodynamic timescale (we choose $U = 1$ as the flow velocity magnitude).
\added{The resulting behavior of $\mathit{St}$ as a function of $\rho_1$ is depicted in figure~\ref{fig:StvsRho1}, showing that it can be well described by a linear law, as expected}. \replaced{}{Such a linear behavior suggests to express the relationship between $\rho_1$ and $\mathit{St}$ in terms of the slender body theory~\mbox{\citep{quennouz2015transport}:}
where $\mu$ is the dynamic viscosity of the fluid and $a = d / c$ is the fiber aspect ratio, and $e$ is the Napier's constant.}
\added{Specifically, the best fit of $\St$ vs $\rho_1$ gives $\St = \alpha \rho_{1}$ with $\alpha \approx 0.04$. }

\added{
Because in the beam equation only one relaxation time is involved in the passive case, its value has been identified as the rotational Stokes time
measured in the active case.}

\subsubsection{\label{sec:Grad}\added{Normal derivative of longitudinal velocity component}}

\begin{figure}
\centering
\includegraphics{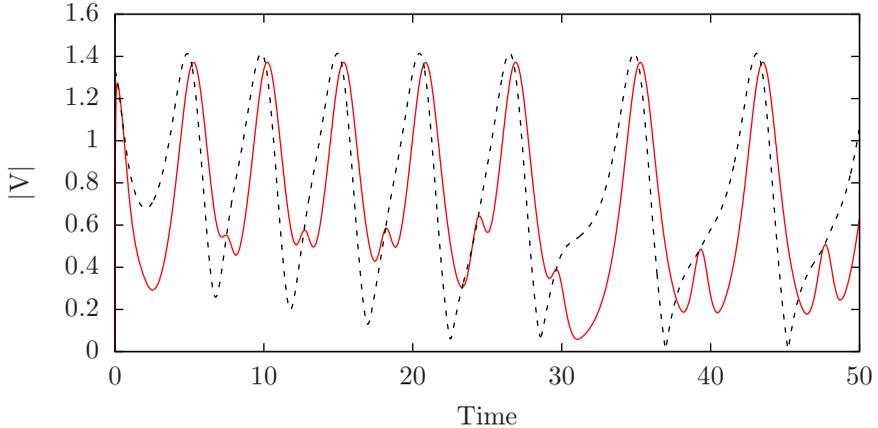}
\caption{Velocity time series at one fiber end for the fiber (red solid line) and for the underlying unperturbed flow (black dashed line), for a fiber with \replaced{$\St \approx 0.01$ }{$\St \approx 0.1$} released into the BC flow, computed with the active model.}\label{fig:velFibEndvsFlow}
\end{figure}

We are now ready to investigate the capability of a rigid fiber to act as a proxy of a laminar, cellular flow in terms of a few fiber properties such as its position and the velocity of the fiber end points.
Figure~\ref{fig:velFibEndvsFlow} reports the velocity magnitude of one fiber end in time, compared with the velocity magnitude of the unperturbed flow (i.e., in the absence of the fiber) evaluated at the same point.
It is evident that the two quantities differ appreciably. This result indicates that a fiber cannot be used to measure single-point flow quantities as done, e.g., in PIV techniques using tracer particles~\citep{adrian1991particle}.

\begin{figure}
\centering
\includegraphics{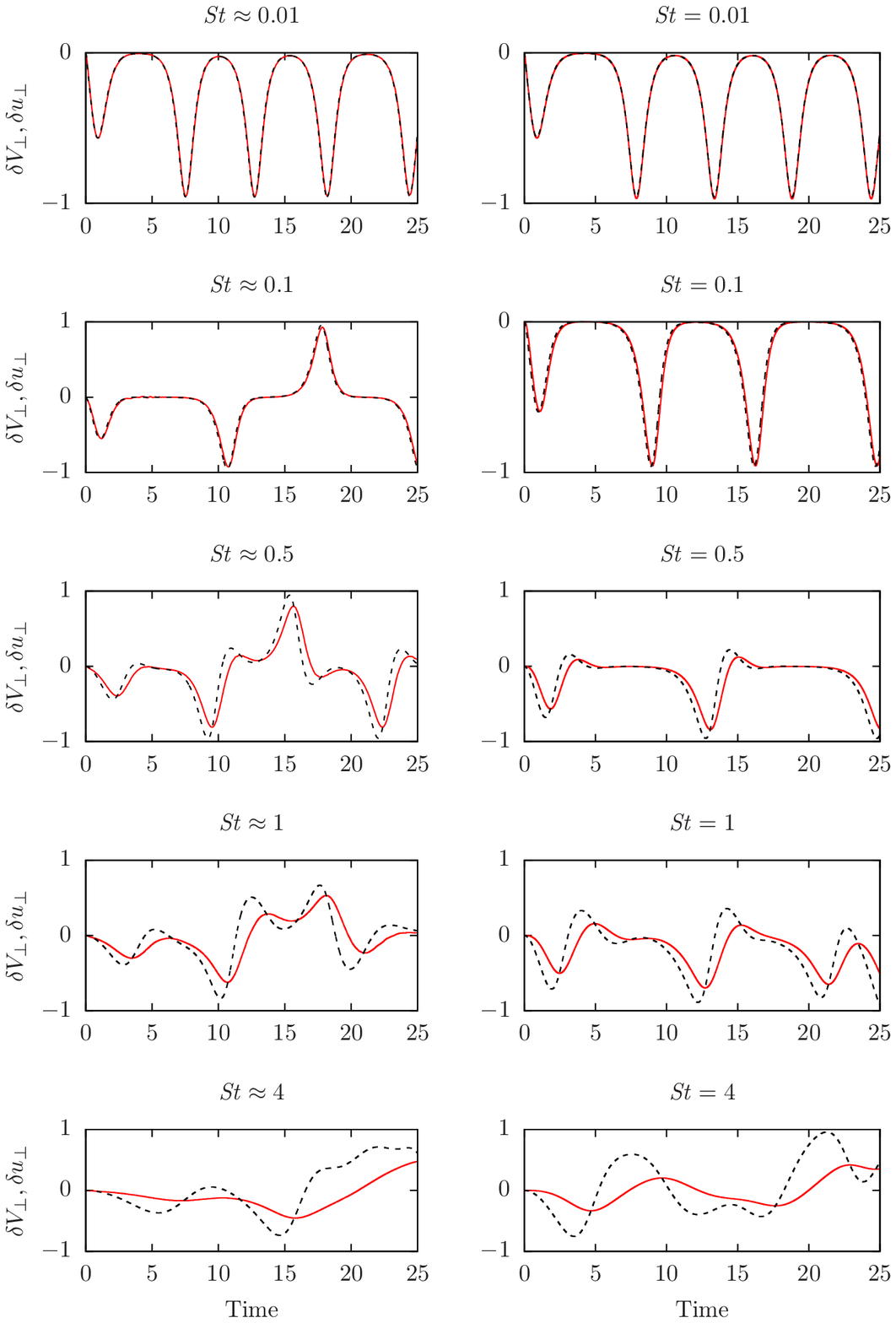}
\caption{Transverse velocity differences of fiber $\delta {V}_\perp$ (solid line) and unperturbed BC flow $\delta {u}_\perp$ (dashed line) for different Stokes number, both for the active (left panels) and passive (right panels) model. The normalized root-mean-square deviation between $\delta {V}_\perp$ and $\delta {u}_\perp$, for both active and passive model, is less than 1\% for $St \leq 0.1$, of the order of 10\% for $St \approx 0.5$ and larger than 15\% for $St \geq 1$. The fiber length is unity and it is thus about $32 \%$ of the size
of the single cell in the considered flow.}\label{fig:DvsSt_BC}
\end{figure}

\added{Hence, we} pass to consider the velocity difference between the fiber end points (figure~\ref{fig:sketch}b), i.e. $\delta \mathbf{V} = \mathbf{V}_B - \mathbf{V}_A$, and denote by $\delta \ub = \ub|_B - \ub|_A$ the corresponding \added{unperturbed} flow velocity difference. 
Comparing directly these two quantities, however, would still yield the same mismatch previously found for the velocity of one end.
Such a mismatch is due not only to the fiber inertia, but also to the fiber inextensibility constraint.
Indeed, if we consider the projection of the velocity difference introduced above along the direction parallel to the end-to-end distance, $\hat{\mathbf{r}}$, for a rigid and inextensible object, such quantity is always zero, although the same quantity for the underlying \added{unperturbed} flow is clearly not.
Our idea is then to project $\delta \mathbf{V}$ on a plane normal to $\hat{\mathbf{r}}$ by simply arguing that along that direction the effect of the inextensibility constraint should be washed out. In terms of the normal unit vector $\hat{\mathbf{r}}_\perp$ (shown in figure~\ref{fig:sketch}b)\added{,} we define the projections:

\begin{equation}
    \delta {V}_\perp = \delta \mathbf{V} \cdot \hat{\mathbf{r}}_\perp,
    \label{eq:dVp}
\end{equation}
\begin{equation}
    \delta {u}_\perp = \delta \mathbf{u} \cdot \hat{\mathbf{r}}_\perp.
    \label{eq:dup}
\end{equation}

The projected quantities~(\ref{eq:dVp}) and~(\ref{eq:dup}) are compared in figure~\ref{fig:DvsSt_BC} where we report the results of our analysis in the BC flow configuration while varying $\St$, for both active and passive models.
For relatively low $\St$ (i.e., the first two rows of the figure), we now notice a remarkable agreement, i.e. the fiber is able to accurately measure the (unperturbed) flow  transverse velocity increments in terms of its transverse velocity increments.
For increasing $\St$, i.e. the fiber inertia, the agreement gets worse, as expected. 
Overall, we observe a close resemblance between results for fibers starting from the same position, using the active (left panels) and passive (right panels) model, especially for the smallest $\St$. 
This suggests that here the effective coupling between the flow and the fiber could be actually neglected, in relation to the measurement of transverse velocity differences. \replaced{}{This is consistent with the strategy we followed to determine the effective value of $a$ via Eq.~(\ref{eq:fit}).} Accordingly, when extending our analysis to three-dimensional and unsteady flows (Sec.~\ref{sec:3D_and_unsteady}) we will exclusively employ the passive solver on the strength of such evidence.

Nevertheless, figure~\ref{fig:DvsSt_BC} tells us that the role of feedback cannot be entirely neglected. For sufficiently large Stokes numbers ($\St \approx 0.1$), the curves on the left panels and those on the right ones are different, \replaced{revealing}{signaling} that the motion of the fiber center of mass is affected by the feedback of the fiber to the flow.
The effect of the latter is indeed crucial when the fiber center of mass moves close to the flow separatrix, potentially causing totally different \deleted{center of mass} trajectories compared to the passive case.


Some further comments are worth to be considered.
First, the projection along the normal direction to the fiber is crucial for the fiber to be a proxy of the flow velocity differences: if we project the velocity differences along a generic direction, the agreement shown before is no longer present (not shown here). 
In the two-dimensional case, the normal direction is uniquely defined by $\hat{\mathbf{r}}_\perp = \pm (\hat{r}_2, -\hat{r}_1,0)$.
In the three-dimensional case we have instead an infinite number of directions belonging to the normal plane to the fiber orientation. We retain $\hat{\mathbf{r}}_\perp = (\hat{r}_2, -\hat{r}_1,0)$ also for 3-D cases (see \S~\ref{sec:3D_and_unsteady}) even if the results do not change for a different choice of $\hat{\mathbf{r}}_\perp$.

The situation considered in figure~\ref{fig:DvsSt_BC} refers to a fiber \added{whose length is about $32 \%$} the size of the single cell. \replaced{This is a case where the fiber length is sufficiently small compared to the variation scale of the flow. Under such condition the velocity difference between the free ends can be compared with the flow gradient evaluated at the fiber center of mass.}{The fiber was thus not evolving in a linear flow.  Let us now focus on the case where
the fiber length is sufficiently small compared to the variation in
space of the flow. In this case, the velocity difference between the free ends can be compared with the flow gradient evaluated at the fiber center of mass.}
For the latter, the same projection along $\hat{\mathbf{r}}_\perp$ has to be applied as before. However, due to the tensorial structure of the gradient $\partial_{j} u_{i}$, this translates to considering a double projection, first along the tangential and then along the normal direction\deleted{to the fiber}: 
\begin{equation}
 D = \partial_{j} u_{i} {\hat{r}_{j}}{\hat{r}_{i}^\perp}. 
 \label{eq:grad}
\end{equation}
In figure~\ref{fig:DvsSt_BC}, where $c / L = ({2 \pi})^{-1}$, the curve representing $D$ is not reported but would be essentially superimposed to that of the fluid velocity difference. \added{Doubling the fiber length, i.e. $c / L = {\pi}^{-1}$, the accuracy gets worse. This simply means that, despite the fact that the fiber accurately measures the transverse velocity differences across the fiber ends, the fiber is too long to allow the derivative to be well approximated by the ratio of the increments.
}

Finally, another aspect to be considered is the tendency of inertial particles to sample preferential zones of the flow, giving rise to peculiar features such as small-scale clustering of dilute suspensions.
This phenomenon is well-known and has been thoroughly investigated for spherical particles in turbulent flows (see, e.g.,~\citet{eaton1994,bec2006,bec2007}) and has also been observed for anisotropic particles (see~\citet{voth2017review} and references therein). 
While this mechanism could impact on the potential of using fibers as a proxy of the flow (i.e., measuring only certain regions of space), we highlight that decreasing the Stokes number, along with improving the measure in itself, assures at the same time that preferential sampling is reduced.

\begin{figure} 
\centering
\advance\leftskip-0.9cm
\includegraphics{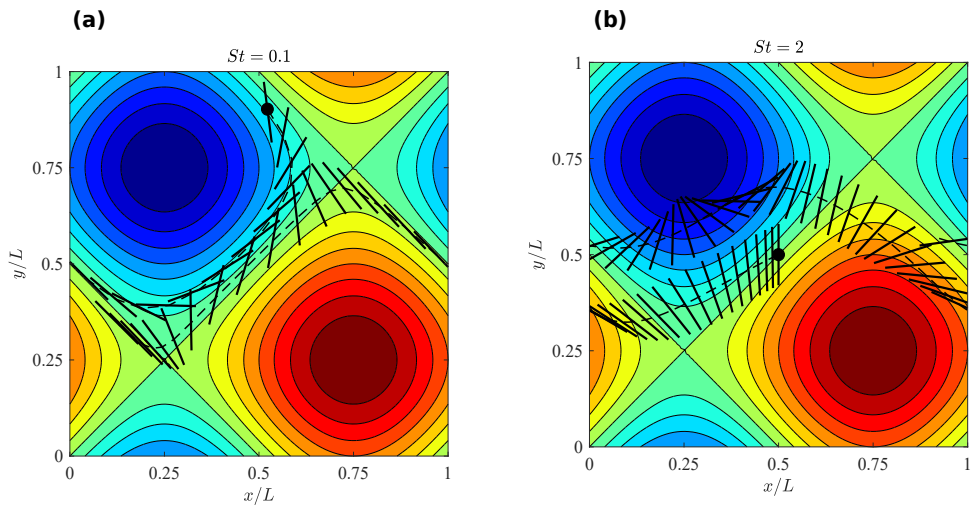}
\caption{Superposition of fiber positions at different instants within the periodic two-dimensional BC flow (the colormap denoting the stream function), computed with the active model. (a) $\St = 0.1$ and (b) $\St = 2$. The dashed line represents the trajectory of fiber center of mass and the black circle indicates its starting position.}
\label{fig:traj}
\end{figure}

In steady cellular flows, we can easily observe the role of inertia in this regard by looking at the fiber trajectories and, in particular, their deviation from the flow streamlines. 
A visualization from our data is given in figure~\ref{fig:traj} where we compare two cases at different $\St$: the less inertial fiber with $\St = 0.1$ initially follows the cell streamlines but, due to the centrifugal effect, eventually reaches the flow separation lines (figure~\ref{fig:traj}a), while the heavier fiber with $\St = 2$ shows a stronger deviation of its trajectory from the streamlines, resulting in a more diffusive behavior (figure~\ref{fig:traj}b).

\begin{figure}
\centering
\includegraphics{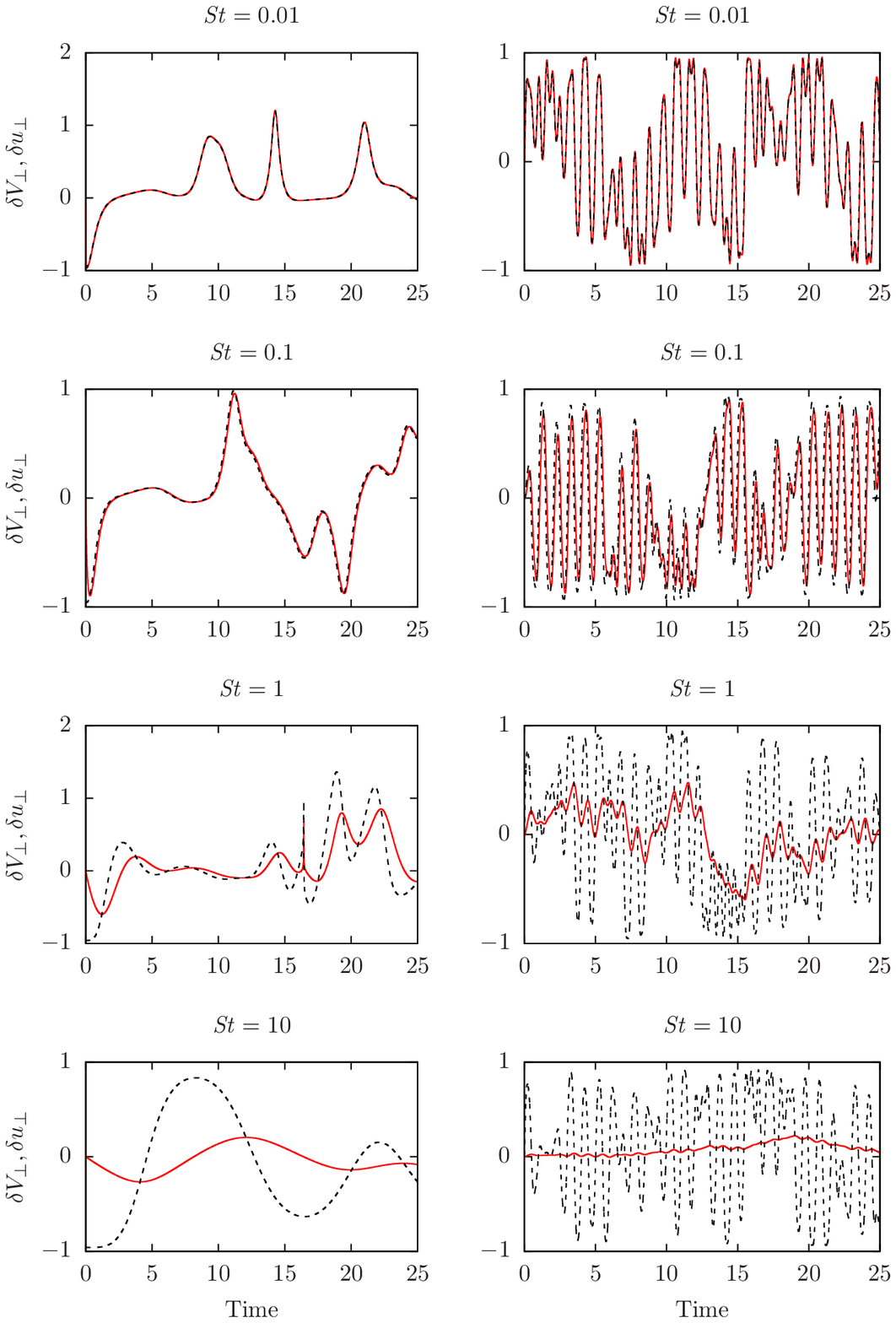}
\caption{Transverse velocity differences of fiber $\delta {V}_\perp$ (solid line) and unperturbed flow $\delta {u}_\perp$ (dashed line) for different Stokes numbers, computed using the passive model. Left panels: ABC flow (Eq.~\ref{eq:ABC}); right panels: 2-D oscillating cellular flow (Eq.~\ref{eq:BCOsc}. \added{The normalized root-mean-square deviation between $\delta {V}_\perp$ and $\delta {u}_\perp$, for both ABC flow and 2-D oscillating flow, is less than 1\% for $St \leq 0.1$, of the order of 15\% for $St = 1$ and larger than 20\% for $St = 10$}.
}\label{fig:DvsSt_ABC}
\end{figure}


\subsection{Extension to three-dimensional and unsteady flows}
\label{sec:3D_and_unsteady}

As a further step, we test the capability of the fiber to measure the transverse velocity differences in three-dimensional steady or unsteady cellular flows.
In light of our findings for the steady BC flow (Sec.~\ref{sec:BC}), we present only results obtained with the passive model, although checks using the active model have been performed and yield the same overall scenario as in the steady, two-dimensional case.

First, we consider the so-called Arnold-Beltrami-Childress (ABC) flow, which is known to be a time-independent, three-dimensional solution of Euler's equations~\citep{dombre1986chaotic}:

\begin{equation}
 \begin{array}{ll}
    u &= \sin z + \cos y \\
    v &= \sin x + \cos z \\
    w &= \sin y + \cos x
  \end{array}
  \label{eq:ABC}
\end{equation}

Unlike the two-dimensional BC flow previously considered, in this flow configuration the Lagrangian fluid elements show both regular and chaotic trajectories, depending on their initial position \mbox{\citep{biferale1995eddy}}. A detailed analysis of this dynamical system can be found in~\mbox{\citet{dombre1986chaotic}}. 

In figure\mbox{~\ref{fig:DvsSt_ABC} (left panels)} we present the results of our analysis for this 3-D case. 
As for the BC flow, we find that for sufficiently low Stokes numbers, i.e. $\St \leq 0.1$, the agreement is evident between the fluid and fiber transverse velocity differences. Like the 2-D case and as expected, the agreement deteriorates for increasing $\St$.

Next, we present the results for the unsteady, i.e. time-periodic, and two-dimensional flow:
\begin{equation}
 \begin{array}{ll} 
    u &= \sin[x+\epsilon_{1}\sin(\omega_{1}t)]\cos[y+\epsilon_{2}\sin(\omega_{2}t)]  \\
    v &= -\cos[x+\epsilon_{1}\sin(\omega_{1}t)]\sin[y+\epsilon_{2}sin(\omega_{2}t)]
 \end{array}
  \label{eq:BCOsc}
\end{equation}\\
where $\epsilon_{1} = \epsilon_{2} = 0.2 L$ are the amplitudes while $\omega_{1} = 2\pi$ and $\omega_{2}=1$ are the frequencies of the oscillation along $x$ and $y$, respectively.
This choice corresponds to a situation where the Lagrangian trajectories of fluid particles are chaotic~\citep{cartwright2010chaotic, Castiglione_1998}.
The projected velocity differences are shown in figure~\ref{fig:DvsSt_ABC} (right panels), where the same conclusions drawn for steady configurations are confirmed: the agreement between the fiber-based measurement and the direct evaluation using the flow expression~(\ref{eq:BCOsc}) increases by decreasing the rotational Stokes number.

\begin{figure} 
\centering
 \includegraphics{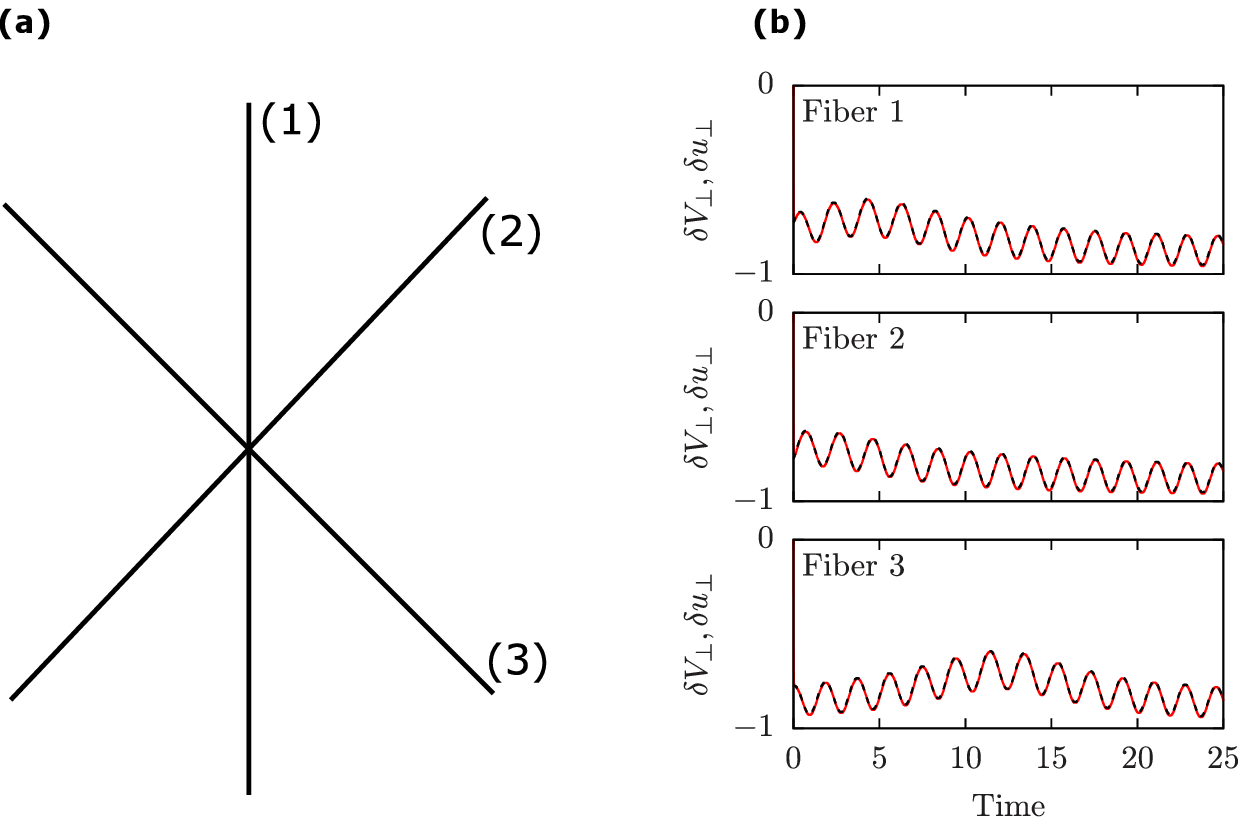}
\caption{(a) Sketch of fiber assembly; (b) transverse velocity differences of each fiber $\delta {V}_\perp$ (solid line) and underlying unperturbed flow $\delta {u}_\perp$ (dashed line) in the case of BC flow with \added{$c = 1$ and} $\St \approx 0.01$, computed using the active model.}
\label{fig:dv3fib}
\end{figure}

\subsection{Evaluation of the velocity gradient tensor}
\label{sec:gradient}


Having characterized the behavior of single fibers, we can move further, focusing on how to access the full velocity gradient tensor $\partial_{j} u_{i}$ and not only its normal-directional projection. This will be achieved by  assembling several fibers in a proper way and exploiting the following idea: for each fiber, Eq.~(\ref{eq:grad}) holds, where the velocity gradient becomes the unknown variable if we use $\delta V_\perp$ (that is measured by tracking the fiber trajectory) in place of $D$. Considering an assembly made by $N_\mathrm{f}$ fibers, we thus have a system of $N_\mathrm{f}$ equations, from which the gradient can be obtained. 

Let us therefore estimate the number of fibers that are needed in the two-dimensional case: here $\partial_{j} u_{i}$ is made by $2 \times 2$ elements; however, the number of independent quantities is reduced of one by exploiting incompressibility. Hence, the assembly has to be made by $N_\mathrm{f} = 3$ fibers, yielding the following system to solve:
\begin{equation}
\begin{array}{ll}
   \delta {V}_\perp^{(1)} = \partial_{j} u_{i} {\hat{r}^{(1)}_{j}}{\hat{r}_{i}^{\perp(1)}}\\ \\
    \delta {V}_\perp^{(2)} = \partial_{j} u_{i} {\hat{r}_{j}^{(2)}}{\hat{r}_{i}^{\perp(2)}}\\ \\
      \delta {V}_\perp^{(3)} = \partial_{j} u_{i} {\hat{r}_{j}^{(3)}}{\hat{r}_{i}^{\perp(3)}}
\end{array}
\label{eq:systm}
\end{equation}
In this system the final number of unknowns are three (i.e.  three components of $\partial_j u_i$ out of four because of the incompressibility condition). Both the left-hand side of the equations and the coefficients of the velocity derivative tensor are easily measurable at each time step along the fiber trajectories and thus known from the numerical experiments. The system can be thus easily solved at each time step while following the fiber along its trajectory.

The three fibers will be connected at their centroids (numerically, it is convenient to realize these connections using springs with sufficiently high stiffness so that the distance between centroids results negligible). 
However, we shall let each fiber to behave as in the single case, its dynamics not being substantially altered by the link with the others. To this end, it is crucial to avoid any rotational restraint, so that fibers are able to rotate freely with respect to each other. \added{For the assembly, we measured the rotational Stokes number following the same procedure described in Sec.~\ref{sec:StNum} for the single fiber. The resulting relaxation times of each fiber composing the assembly turned out to be the same as the rotational Stokes time of the single  isolated fiber. One can thus conclude that the Stokes time of the assembly is the same of a single fiber.} 

\begin{figure}
\centering
\includegraphics{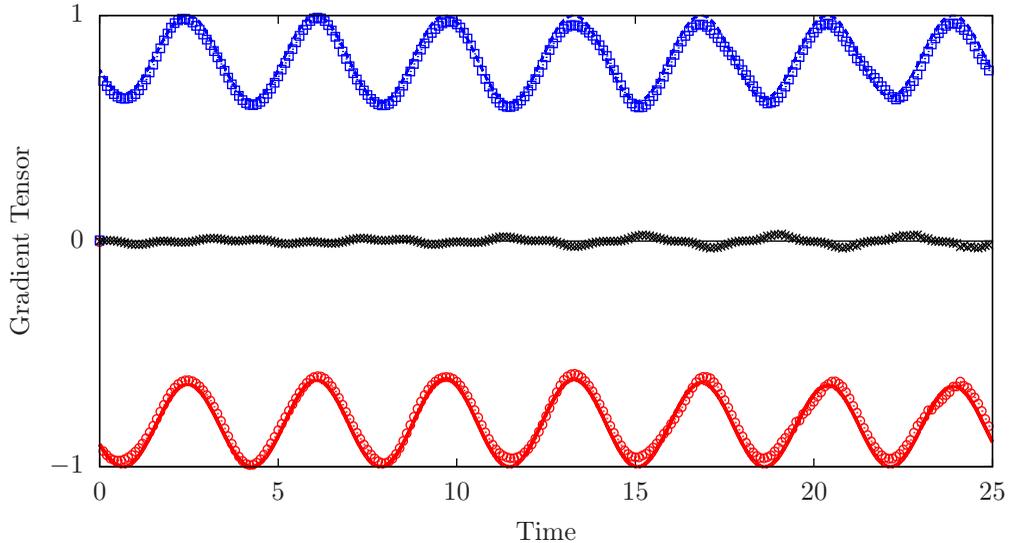}
\caption{Velocity gradient tensor components in the BC flow (\ref{eq:BC}) reconstructed by tracking the fiber assembly with $\St \approx 0.01$, using the active model. Red circles: $\partial_y u$ obtained by the fiber Lagrangian tracking; red solid line: $\partial_y u$ for the unperturbed flow; blue squares: $\partial_x v$ obtained by the fiber  Lagrangian tracking; blue dashed line $\partial_x v$ for the unperturbed flow; black crosses: $\partial_x u = -\partial_y v$ obtained by the fiber  Lagrangian tracking; black line: $\partial_x u = -\partial_y v = 0$ for the unperturbed flow. \added{The normalized root-mean-square error between the components of the gradient tensor reconstructed by the Lagrangian tracking and those of the unperturbed flow is of the order of 1\%.}}
\label{fig:grad_att}
\end{figure}

\replaced{Now we are ready to test the outlined concept}{The outlined concept is tested} in the steady BC flow already used in Sec.~\ref{sec:BC}. As a first step, we look at the resulting time histories of the projected velocity difference for each fiber composing the assembly (figure~\ref{fig:dv3fib}), recovering the same evidence found in the case of single fibers. This provides a clue that also in this configuration it is possible to capture the features of the fluid flow. 
Indeed, we proceed to combine the information from all fibers, finally obtaining the velocity gradient tensor as shown in figure~\ref{fig:grad_att}, where the time series of each element of $\partial_j u_i$ is reported, both for the fiber Lagrangian tracking and for the corresponding analytical value of the unperturbed flow. The comparison between the two quantities yields good agreement, with differences that are ascribed to numerical resolution and the finite inertia of fibers.

\begin{figure} 
\centering
\includegraphics{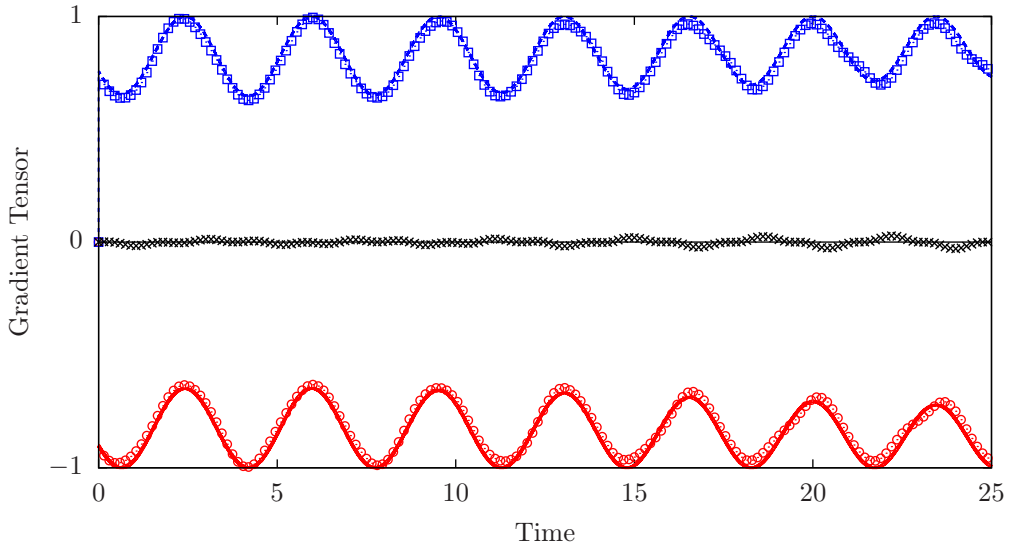}
\caption{As in figure~\ref{fig:grad_att} but for assembly of passive fibers. \added{The normalized root-mean-square error between the components of the gradient tensor reconstructed by the Lagrangian tracking and those of the unperturbed flow is of the order of 1\%.}}
\label{fig:grad_pass}
\end{figure}

\begin{figure} 

\includegraphics{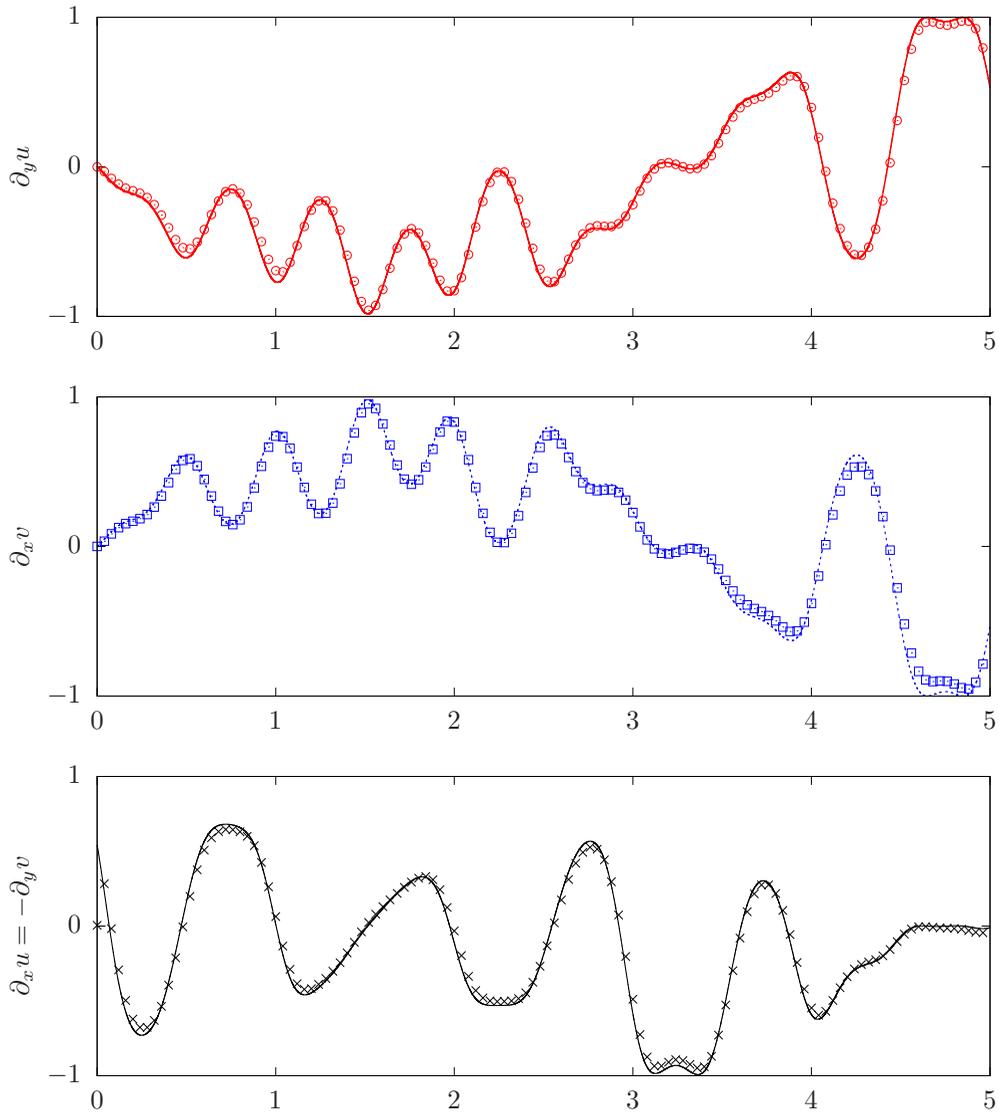}
\caption{As in figure~\ref{fig:grad_pass} but for the oscillating two dimensional flow (\ref{eq:BCOsc}) (the figure is split into three panels for a better readability of the gradient tensor components). \added{The normalized root-mean-square error between the components of the gradient tensor reconstructed by the Lagrangian tracking and those of the unperturbed flow is of the order of 1\%.}}
\label{fig:grad_Oscpass}
\end{figure}

The reported results are for the active model but closely similar evidence is obtained using the passive model. In figure~\ref{fig:grad_pass} we show the result for the assembly of passive fibers in the static BC flow, highlighting essentially the same behavior obtained in the active case (note that the same initial condition for the assembly of fibers was used in both cases).
Finally, we complement the analysis by employing an assembly of passive fibers in the oscillating two dimensional flow introduced in Sec.~\ref{sec:3D_and_unsteady} (Eq.~\ref{eq:BCOsc}). Results are shown in figure~\ref{fig:grad_Oscpass} from which we can confirm the same conclusion as outlined before.

\begin{figure} 
\centering
\includegraphics{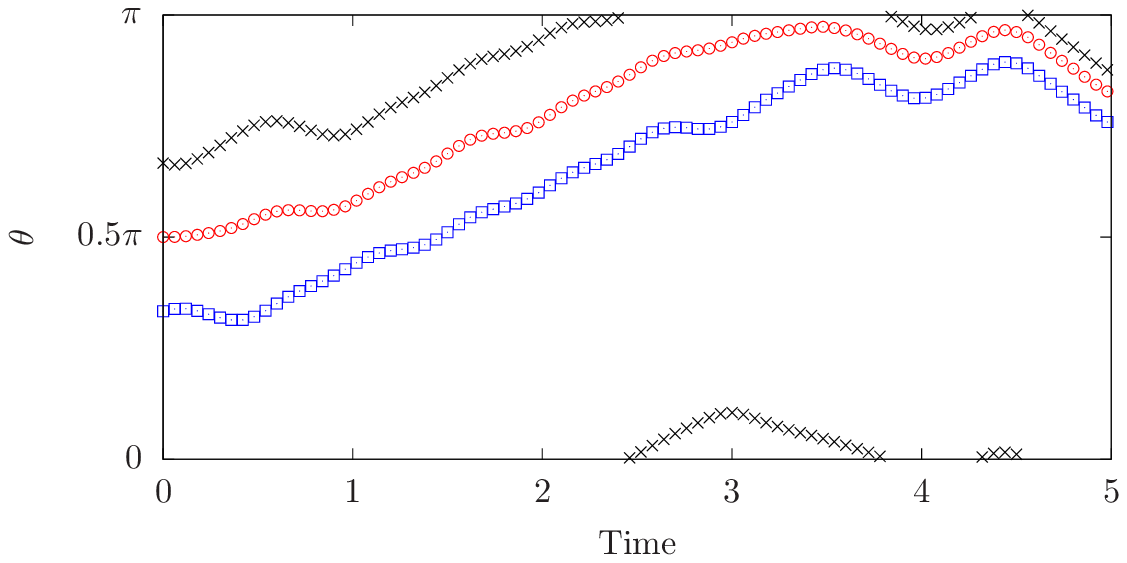}
\caption{Time history of the angles (rad) of the three fibers composing the assembly in the two dimensional oscillating flow. Red circles: fiber 1; blue squares: fiber 2; black crosses: fiber 3. Angles are measured with respect to the horizontal direction. The minimum value of the standard deviation is $0.18$.}
\label{fig:angoli}
\end{figure}

\added{It is worth noting that the linear system~(\ref{eq:systm}) we numerically solved to obtain all components of the flow gradient can become overdetermined because of the alignment of two of more fibers.
To avoid this problem we found that the simple recipe to impose a small displacement (of $\Delta s$, the size of the Lagrangian mesh) between the centroids of the fibers of the assembly (instead of imposing them to be zero) is enough to prevent perfect alignment of the fibers, thus preventing the breakdown of the solution.
The results reported in figure~\ref{fig:grad_att},~\ref{fig:grad_pass} and~\ref{fig:grad_Oscpass} have been obtained exploiting this simple, but effective, strategy.
Figure~\ref{fig:angoli} reports the time history of the angle of the three fibers composing the assembly, in the oscillating two dimensional flow. As shown in figure, in the considered time frame the alignment between the fibers does not occur and the minimum value of the standard deviation between the three angles is $0.18$ rad. Extending the time frame (not shown), up to 25, the minimum value of the standard deviation we measured was $0.0085$ rad which was however large enough to accurately solve the system.}

\section{\label{sec:discussion}Conclusions and perspectives}
This study focused on the capability of measuring the whole structure of the velocity gradient   in steady, unsteady regular and chaotic cellular flows by means of Lagrangian tracking of assembly of rigid fibers. 
Two different kinds of fiber models have been considered: a fully-coupled fiber described in terms of an immersed-boundary method and a passive fiber described by the slender body theory.
We first characterized the role of fiber inertia by defining a rotational Stokes number, which is evaluated as a function of \deleted{other parameters such as} the fiber linear density.
Hence, considering the velocity difference between the fiber end-points and the same difference concerning the underlying fluid velocity, both projected along the normal direction to the fiber, \added{we showed that} the fiber turns out to be a proxy of such two-point quantity.
For sufficiently small fibers, this two-point quantity reduces to the transverse component of the flow velocity derivative along the fiber direction.
Furthermore, the comparison between results obtained for the active model and the passive model suggests that the coupling between the flow and the fiber could be neglected, at least for small $\St$.

This capability of \added{using rigid fibers as a way of measuring flow properties} has potential application in experimental \deleted{measurement} techniques allowing to access small-scale, multi-point properties of fluid flows, offering an alternative to other methods that have been proposed which rely on complex elaborations using PIV/PTV~\citep{hoyer2005,krug2014,lawson2014}. Future work will thus be devoted to the practical implementation of the outlined concept in a laboratory environment. Preliminary results in this direction\added{ considering rigid fibers of millimetric size dispersed in turbulent flow appear very encouraging~{\citep{Brizzo2019}},} confirming the validity of our idea \deleted{of using rigid fibers as a way of measuring flow properties, an idea that seems successful} \added{in a framework} well beyond the laminar/chaotic examples analyzed here. The \added{applicability of the concept} can \added{thus} be extended to \deleted{measurements of} three-dimensional and/or turbulent flows, along with considering assemblies of fibers that would be able to \added{measure} the full structure of the velocity gradient.

\section*{Acknowledgements}
The authors warmly acknowledge A.~A. Banaei, M.~E. Rosti and L.~Brandt (KTH, Sweden) for sharing an initial version of the code for the active solver.
CINECA and INFN are also acknowledged for the availability of high performance computing resources and support. AM thanks the financial support from the Compagnia San Paolo, project MINIERA n. I34I20000380007.

\section*{Declaration of Interests}
The authors report no conflict of interest.

\appendix
\section{Immersed boundary method}
\label{sec:appendix_IBM}

This appendix presents the numerical procedure regarding the active approach introduced in Sec.~\ref{sec:active}.
The Navier-Stokes equations~(\ref{eq:NS1}) and~(\ref{eq:NS2}) are solved for a cubic domain of side $L$ with periodic boundary conditions in all directions, which is discretized into a Cartesian grid using $N$ points per side. The solution is obtained by using a finite difference, fractional step method on a staggered grid with fully explicit space discretization and third-order Runge-Kutta scheme for advancement in time.
Finally, the resulting Poisson equation enforcing incompressibility is solved using Fast Fourier Transform.

As for the fiber-flow interaction, we employ the IB approach of~\citet{huang_shin_sung_2007a} and later modified by~\citet{banaei2019numerical}. The Lagrangian forcing is first evaluated at each fiber point, in order to enforce the no-slip condition $\dot\Xb=\Ub(\Xb(s,t),t)$, as
\begin{equation}
    \Fb(s,t) = \beta \, (\dot{\Xb} - \Ub),
\end{equation}
where $\beta$ \added{is a large negative constant~{\citep{huang_shin_sung_2007a}}} and 
\begin{equation}
 \Ub(\Xb(s,t),t) = \int \ub(\xb,t)\delta(\xb-\Xb(s,t))\,\dd\xb
\end{equation}
 is the interpolated fluid velocity at the Lagrangian point.
A spreading is thus performed over the surrounding Eulerian points, yielding the volumetric forcing acting on the flow
\begin{equation}
   \fb(\xb,t) = \int \Fb(s,t)\delta(\xb-\Xb(s,t))\, \dd s.
\end{equation}
Both the interpolation and spreading feature the Dirac operator, which in discretization terms is transposed into the use of regularized $\delta$; in our case, we employ the function proposed by~\citet{roma1999}.

The described procedure has been implemented and extensively validated in both laminar and turbulent flow conditions: for related information, the reader is referred to~\citet{rosti_brandt_2017a,rosti_izbassarov_tammisola_hormozi_brandt_2018a,shahmardi_zade_ardekani_poole_lundell_rosti_brandt_2019a,rosti2019flowing,banaei2019numerical}.

\bibliographystyle{jfm}
\bibliography{bib}

\end{document}